\documentclass[twocolumn]{jpsj3_660}
\usepackage{txfonts}
\title{Irreversible Heating Measurement with Microsecond Pulse Magnet: \\Example of the $\alpha$-$\theta$ Phase Transition of Solid Oxygen}

\author{Toshihiro Nomura$^1$\thanks{t.nomura@issp.u-tokyo.ac.jp}, 
Yasuhiro H. Matsuda$^1$\thanks{ymatsuda@issp.u-tokyo.ac.jp}, 
Shojiro Takeyama$^1$, 
and Tatsuo~C.~Kobayashi$^2$}
\inst{$^1$Institute for Solid State Physics, University of Tokyo, 5-1-5 Kashiwa, Chiba 277-8581, Japan \\
$^2$Department of Physics, Okayama University, Okayama 700-8530, Japan} 

\abst{Dissipation inevitably occurs in first-order phase transitions, leading to irreversible heating.
Conversely, the irreversible heating effect may indicate the occurrence of the first-order phase transition.
We measured the temperature change at the magnetic-field-induced $\alpha$-$\theta$ phase transition of solid oxygen.
A significant temperature increase from 13 to 37 K, amounting to 700 J/mol, due to irreversible heating was observed at the first-order phase transition.
We argue that the hysteresis loss of the magnetization curve and the dissipative structural transformation account for the irreversible heating.
The measurement of irreversible heating can be utilized to detect the first-order phase transition in combination with an ultrahigh magnetic fields generated in a time of $\mu$s order.
}

\begin{document}
\maketitle
\thispagestyle{plain}
\section{Introduction}
In general, when a material is magnetized or demagnetized under an adiabatic condition, its temperature changes.
This is called the magnetocaloric effect (MCE).
The MCE has potential application for the magnetic refrigeration and has been widely studied with the aim of achieving refrigeration with higher efficiency \cite{Magcal2009CondM,Magcal2014NMat}.
In addition, MCE measurement is recognized as a powerful tool for understanding the thermodynamical properties of magnetic materials \cite{02PRL_URu2Si2,09PRL_YbRhSi,10PNAS_QCP,11Sci_SrRuO,06PRL_URu2Si2,09PRBR_Ba3Cr2O8,12PRL_BiCu2PO6}.
Recently, MCE measurements have been combined with pulse magnetic fields and applied to various research in physics at high fields \cite{Kohama2010RSI,Dresden2012MST,Kihara2013RSI,Kihara2014PRB,Kohama2014PRB_R,Zavareh2015APL}.

In the adiabatic MCE, the temperature change $\Delta T$ can be divided into reversible and irreversible terms \cite{Bates55_separate,Ortin88,Shamberger09} as
\begin{equation}
\Delta T=\Delta T_\mathrm{rev}+\Delta T_\mathrm{irr}.
\end{equation}
The reversible term $\Delta T_\mathrm{rev}$ is due to the change in entropy (typically spin entropy) under a quasi-static condition.
For example, when a paramagnetic material is magnetized (demagnetized), the spin entropy decreases (increases) and the temperature increases (decreases) to conserve the total entropy.
The latent heat of the phase transition is also regarded as this term.
The absorption and evolution of the same amount of heat occur when a parameter goes back and forth between two phases in the phase diagram.

On the other hand, the irreversible term $\Delta T_\mathrm{irr}$ ($>0$) originates from dissipative processes such as the dissipative motion of magnetic domain walls \cite{35PR_Ni,Chikazumi,Bertotti} or superconducting vortices \cite{NBCO05irrev}.
More generally, a dissipative process is inevitably involved in the first-order phase transition because of the potential barrier between two phases.
In these examples, the magnetization curve also shows irreversibility, namely, hysteresis appears.
The hysteresis of the magnetization curve indicates the discordance of the magnetizing energy between up and down sweeps of the magnetic field.
The surplus energy can be obtained from $\mu_0 \oint B dM$, which mainly results in the heating as hysteresis loss \cite{Chikazumi,Bertotti}.

In previous studies 
\cite{02PRL_URu2Si2,09PRL_YbRhSi,10PNAS_QCP,11Sci_SrRuO,
06PRL_URu2Si2,09PRBR_Ba3Cr2O8,12PRL_BiCu2PO6,Kohama2010RSI,Dresden2012MST,
Kihara2013RSI,Kihara2014PRB,Kohama2014PRB_R,Zavareh2015APL}, 
the authors were mainly interested in $\Delta T_\mathrm{rev}$, which reveals isentropes in the magnetic field-temperature ($B$-$T$) plane.
The discontinuity in the isentropes indicates the phase boundary in the $B$-$T$ phase diagram.
In this study, however, we focus on $\Delta T_\mathrm{irr}$.
Irreversible heating is related to dissipation, possibly the first-order phase transition.
Therefore, information on the irreversible heating can be used to detect the first-order phase transition \cite{06PRL_URu2Si2,09PRBR_Ba3Cr2O8,12PRL_BiCu2PO6}.

A pulse magnetic field is suitable for the measurement of $\Delta T_\mathrm{irr}$.
The adiabatic condition is easily realized in a short duration of the field.
In addition, faster pulse magnetic fields result in larger hysteresis loops enhancing the hysteresis loss.
For example, the single-turn coil (STC) technique generates pulse fields of over 100 T within 10 $\mu$s in a destructive manner \cite{03LTP_Miura}.
Here, the sweep speed of the magnetic field reaches the order of $10^7$ T/s.
Thermometric measurement has never been conducted with the STC because of experimental difficulties, such as the inductive voltage, eddy current, shock wave, and short duration.
However, if we are interested in $\Delta T_\mathrm{irr}$, all these difficulties can be avoided simply by measuring the temperature ``just after'' the pulse field generation.
If the sample volume is sufficiently large, the irreversible heating can be measured even after the field generation.

In this paper, we report the irreversible heating at a first-order phase transition induced by the STC.
The $\alpha$-$\theta$ phase transition of solid oxygen, which shows significant hysteresis in the magnetization curve \cite{Nomura14prl,Nomura15prb}, was measured.
This paper is organized as follows.
In Sect. 2, the experimental setting of the irreversible heating measurement is described.
Optical spectroscopy is simultaneously conducted to show the validity of the irreversible heating measurement.
In Sect. 3, the results of the irreversible heating and optical spectroscopy are shown and compared with each other.
In Sect. 4, the observed temperature change is quantitatively discussed in terms of the hysteresis loss.
In Sect. 5, as a conclusion, the advantage of this $\Delta T_\mathrm{irr}$ measurement technique is stated.

\section{Experiment}
\begin{figure*}[tbh]
\centering
\includegraphics[width=16.6cm]{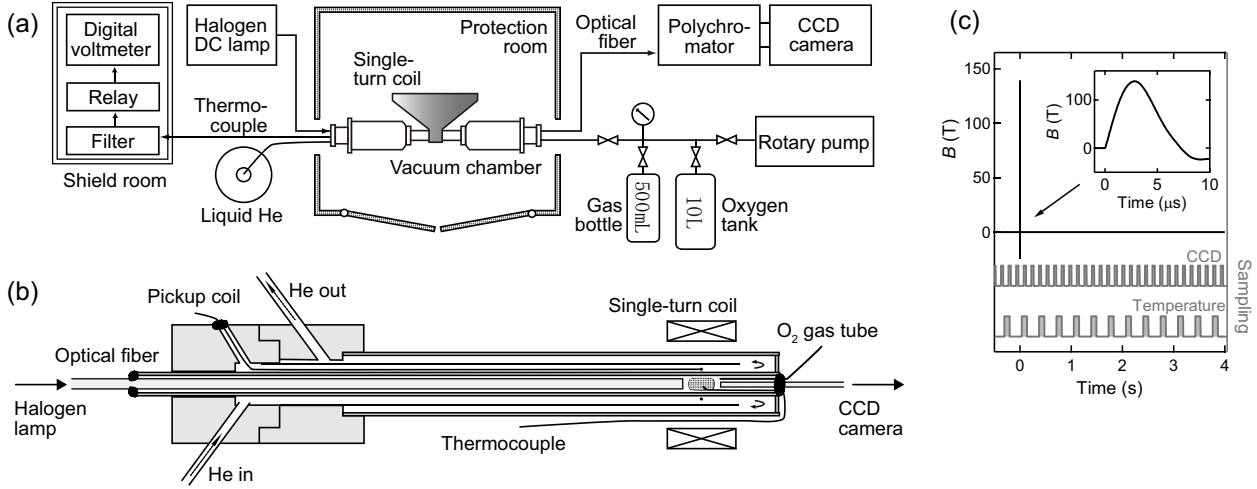}
\caption{\label{fig:system}
(a) Block diagram of the experimental setup for the measurement of the irreversible heating of solid oxygen in ultrahigh magnetic fields.
Optical absorption spectroscopy is simultaneously conducted using the CCD camera.
(b) Enlarged view near the sample space. The vacuum chamber is omitted for simplicity.
(c) Time chart showing the relation between the magnetic field and the sampling rates of the optical spectroscopy and the temperature measurement.}
\end{figure*}

Figure \ref{fig:system}(a) shows a block diagram of the experimental setup.
Magnetic fields of up to 137 T are generated using the STC system in the Institute for Solid State Physics, Univ. of Tokyo \cite{03LTP_Miura}.
The temperature and the optical absorption spectra are simultaneously monitored to detect the irreversible heating.
If irreversible heating occurs, the absorption spectra change owing to the temperature increase.

Figure \ref{fig:system}(b) shows an enlarged view near the sample space.
Solid oxygen is condensed from high-purity O$_2$ gas (99.999\%) in an optical cryostat made of Bakelite as shown by the shadowed area \cite{13LTP_Nomura}.
The sample diameter is 2.4 mm and the optical path length is typically 2 mm.
The sample is sandwiched by two optical fibers for the incident and transmitted light.
The sample temperature is monitored by a chromel-constantan thermocouple (E-type), which is buried directly into the solid oxygen, and therefore the thermal contact is considered to be almost perfect.
The thermoelectric voltage is measured by a digital voltmeter (Keithley 2000).
A low-pass filter and relay are employed to protect the voltmeter from the huge noise of the STC system.
For the optical spectroscopy, a CCD camera and DC halogen lamp are used.
The repetition times of the temperature and CCD measurements are ${\Delta}t_\mathrm{Temp}=0.34$ s and ${\Delta}t_\mathrm{CCD}=0.15$ s, respectively.
A time chart showing the relation between the magnetic field and the sampling rate of each measurement is presented in Fig. \ref{fig:system}(c).
The pulse magnetic field generation instantaneously finishes within a duration of 10 $\mu$s.
The measurements can be regarded as quasi-continuous with a time resolution of 0.34 s.

In the time scale of $\mu$s, magnetization occurs under an adiabatic condition.
Most of the irreversible heating results in the temperature increase of the sample in this time scale.
The thermal relaxation time between the sample and the environment is typically of s order.
For the case of solid oxygen, the thermal relaxation time is estimated as 2 s \cite{relaxation_t}.
Therefore, the irreversible heating can be observed if we measure the temperature within 2 s.
The relaxation time can easily be increased by increasing the sample volume.

\section{Results}
\begin{figure*}[tbh]
\centering
\includegraphics[width=14.2cm]{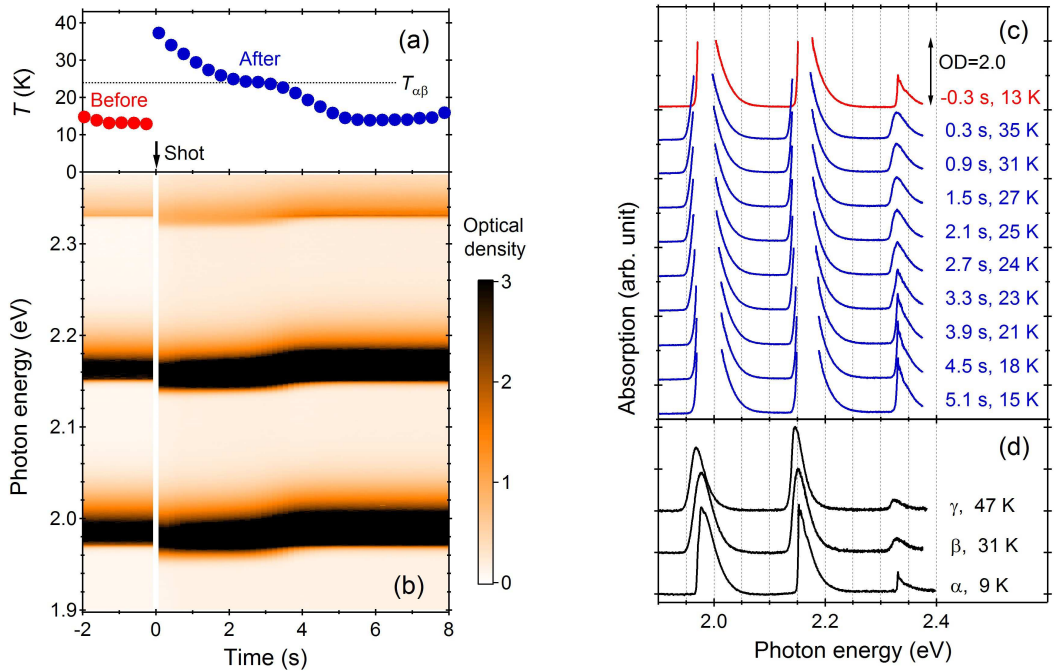}
\caption{\label{fig:Spectra}
(Color online) 
Time dependence of the temperature (a) and optical absorption spectra (b) of solid oxygen.
(c) Absorption spectra at each timing in Fig. \ref{fig:Spectra}(b).
(d) Typical absorption spectra of solid oxygen $\alpha$, $\beta$, and $\gamma$ phases at zero field.
The absorption intensity is normalized by the peak at 2.16 eV.}
\end{figure*}

Figure \ref{fig:Spectra}(a) shows the time evolution of the temperature before and after the pulse field generation.
At 0 s, indicated by the black arrow, a magnetic field of up to 137 T is instantaneously generated and the $\alpha$-$\theta$ phase transition takes place.
The temperature of solid oxygen discontinuously increases from 13 to 37 K at 0 s, and gradually decreases to 15 K.
This is clear evidence of the irreversible heating, originating from the $\alpha$-$\theta$ phase transition of solid oxygen.
The temperature higher than the $\alpha$-$\beta$ transition point ($T_{\alpha\beta}=23.9$ K \cite{04PR_Freiman}) means that the $\beta$ phase appears after the field generation.

Figure \ref{fig:Spectra}(b) shows the time evolution of the absorption spectra.
The optical density, OD $=\mathrm{log}_{10}(I_0/I)$, is mapped in a color scale where $I_0$ and $I$ are the incident and transmitted light intensities, respectively.
The absorption spectra at each timing are shown in Fig. \ref{fig:Spectra}(c).
The dynamic range of the present measurement is up to OD = 2.
The absorption peak at 1.98 eV is due to the bimolecular absorption of solid oxygen, 
${}^3\Sigma{}_g^-{}^3\Sigma{}_g^-$$\rightarrow$${}^1\Delta{}_g{}^1\Delta{}_g$ \cite{bm61_exp_landau,absorptionseries68pss,bm_theory2,76BMAbs_Exp}.
The two peaks at 2.16 and 2.34 eV are vibronic replicas ($v=1,\ 2$).
The absorption spectra of each phase of solid oxygen at zero field are compared in Fig. \ref{fig:Spectra}(d).
The absorption peak shape depends on the magnetic state in these phases \cite{bm61_exp_landau,absorptionseries68pss}.
Therefore, each phase can be distinguished by the shape of the absorption peak.
Here, we focus on the shape of the $v=2$ absorption peak at 2.34 eV.
The absorption peak shape clearly changes at 0 s, which corresponds to the transformation from the $\alpha$ to $\beta$ phase.
This is consistent with the result of the temperature measurement, which indicates that the temperature increases to the range of the $\beta$ phase ($T>T_{\alpha\beta}=23.9$ K).
Therefore, the emergence of the $\beta$ phase is confirmed by optical spectroscopy.
After the $\beta$ phase appears at 0 s, the temperature gradually decreases and passes through $T_{\alpha\beta}$ at 2.7 s.
At the same time, the shape of the absorption spectra changes from that of $\beta$ to that of $\alpha$.
This accordance indicates that thermal equilibrium between the thermocouple and the sample is realized.

An experiment with a lower maximum field is also conducted to confirm that the observed temperature increase is an intrinsic effect originating from the phase transition.
Here, the maximum field strength is set to $B_\mathrm{Max}=103$ T, and the initial temperature measured immediately before the field generation is $T_0=10$ K.
No phase transition occurs under these conditions \cite{Nomura14prl,Nomura15prb}.
In this control experiment, no temperature change is observed within an accuracy of $\pm$1 K.
This finding means that the extrinsic effects, such as the eddy current at the metallic wires or the radiation from the STC explosion, are negligible for the measured temperature.
Therefore, the observed heating effect can be attributed to the phase transition.
These experimental findings mean that the irreversible heating is significant at the $\alpha$-$\theta$ phase transition.

\section{Discussion}
\begin{figure}[tbh]
\centering
\includegraphics[width=6cm]{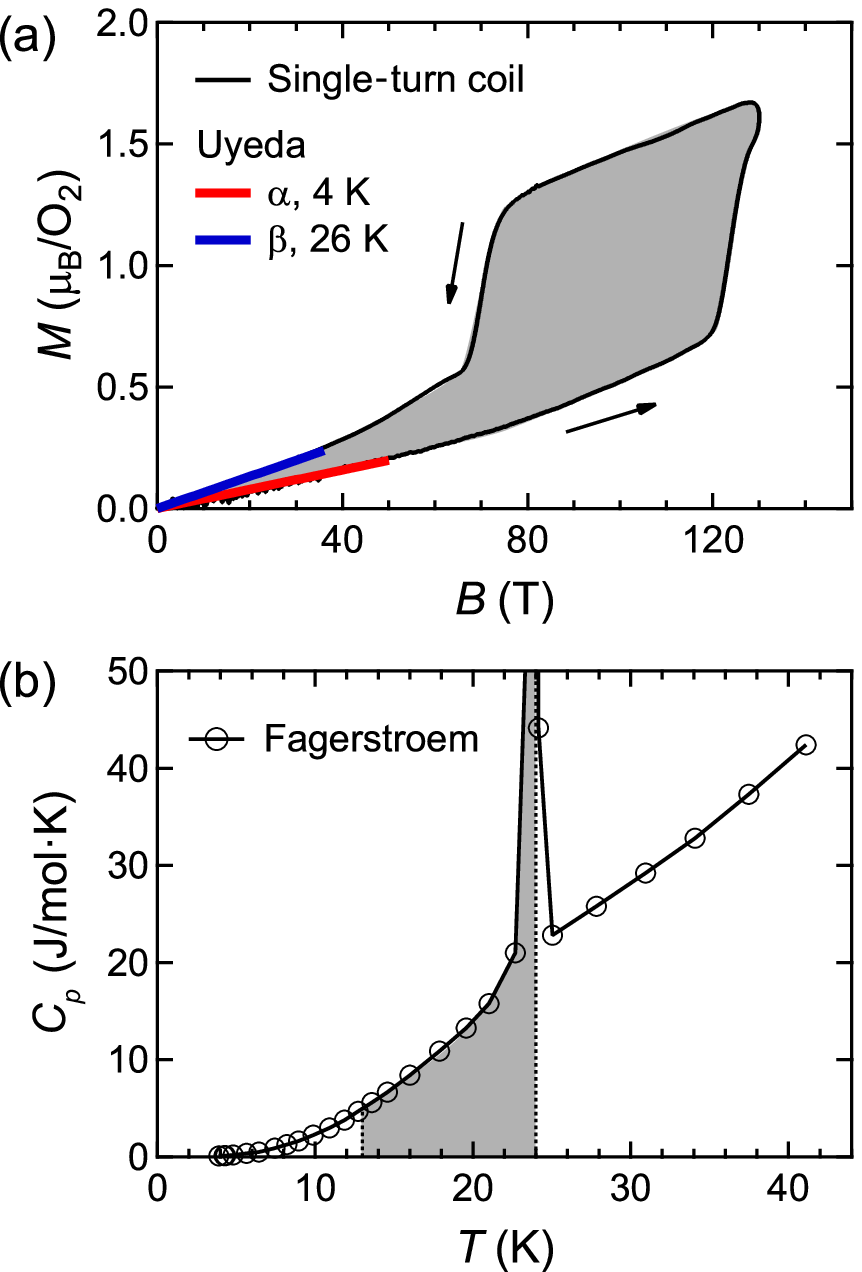}
\caption{\label{fig:MGN_C}
(Color online) 
(a) Magnetization curve of solid oxygen under the conditions of $B_\mathrm{Max}=130$ T and $T_0=10$ K.
Results of the nondestructive magnet reported by Uyeda {\it et al.} are also shown \cite{85JPSJ_Uyeda}.
(b) Specific heat curve reported by Fagerstroem and Hollis Hallett \cite{69LTP}.
In (a) and (b), the gray areas are set to be same (330 J/mol).
The temperature increase is estimated to be from 13 to 24 K.}
\end{figure}

Here, we discuss the amount of irreversible heating $\Delta T_\mathrm{irr}$ in terms of the hysteresis loss of the magnetization curve.
Figure \ref{fig:MGN_C}(a) shows the magnetization curve of solid oxygen measured by using the STC under the conditions of $B_\mathrm{Max}=130$ T and $T_0=10$ K \cite{Nomura14prl,magnetization}.
The initial slope of the magnetization curve is calibrated by a result obtained using a nondestructive pulse magnet \cite{85JPSJ_Uyeda}.
The magnetization jump at 125 T in the up sweep of the field is due to the $\alpha$-$\theta$ phase transition of solid oxygen.
The magnetization curve shows a large hysteresis loop, where the hysteresis loss is obtained as $\mu_0 \oint B dM=330$ J/mol.
In this study, we assume that all of this dissipated energy results in sample heating, and the other dissipation paths such as acoustic emission (Barkhausen effect \cite{Chikazumi,Bertotti}) are neglected.
The temperature increase is estimated to be from 13 to 24 K by equalizing the integrated areas of the heat capacity and hysteresis loss (gray areas in Fig. \ref{fig:MGN_C}).

In the irreversible heating measurement, the temperature increase is observed to be from 13 to 37 K. 
The amount of heat is calculated as $\Delta Q=\int_{13\ \mathrm{K}}^{37\ \mathrm{K}}C_p dT=700$ J/mol using the heat capacity data shown in Fig. \ref{fig:MGN_C}(b) \cite{69LTP}.
This is more than twice the value estimated from the magnetization curve.
This discrepancy cannot be explained by the slight difference in the experimental conditions: $B_\mathrm{Max}=137$ T in the temperature measurement, and $B_\mathrm{Max}=130$ T in the magnetization curve.
It is necessary to take another dissipation mechanism into account.

Here, we propose a structural transformation as the origin of the additional dissipation.
In the above discussion, we only consider the hysteresis loss of the magnetization curve, $\mu_0 \oint B dM$.
However, in the $\alpha$-$\theta$ phase transition, drastic molecular rearrangement occurs at the same time \cite{Nomura14prl,Nomura15prb}.
At the structural transformation, additional energy is consumed to overcome the frictional force opposing the domain boundary motion.
Such dissipation will contribute to the irreversible heating in addition to the magnetic hysteresis loss.
In other words, $700-330=370$ J/mol is proposed to originate from the dissipation of the structural transformation.

The experimental results and the above discussion do not answer the question of at which timing the irreversible heating occurs.
For the case of ferromagnets \cite{Bates55_separate,35PR_Ni}, the irreversible heating mainly occurs at the magnetization jump, namely when the domain reorientation occurs.
For other magnetic materials \cite{06PRL_URu2Si2,09PRBR_Ba3Cr2O8,12PRL_BiCu2PO6}, it occurs at the first-order phase transition.
Therefore, in the same way as for solid oxygen, the irreversible heating is considered to occur at the $\alpha$-$\theta$ phase transition with magnetization jumps in the up and down sweeps.
Because of this heating, the phase transformation occurs as $\alpha \rightarrow \theta$ in the up sweep, while $\theta \rightarrow \beta$ in the down sweep.
This is confirmed by the magnetization curve in Fig. \ref{fig:MGN_C}(a); the magnetization value in the down sweep coincides with that of the $\beta$ phase.
The absorption spectra reported in Ref. 24 also show the same behavior.

Finally, we discuss the possibility of other extrinsic origins of the irreversible heating that may be confusing with the first-order phase transition.
For example, if the sample is conductive, the eddy current induced by pulsed magnetic fields results in irreversible heating.
Generally, the delayed response of the magnetization results in hysteresis.
Therefore, the spin glass state or superparamagnetism can also result in hysteresis if the relaxation time of the dynamics is comparable to the sweep speed of the pulse magnetic field \cite{Chikazumi}.
To eliminate these other possibilities, the $B_\mathrm{Max}$ dependence of the irreversible heating needs to be examined.
If the irreversible heating originates from the first-order phase transition, the $B_\mathrm{Max}$ dependence will change discontinuously at the critical field.
For the other origins, the heating continuously increases as $B_\mathrm{Max}$ increases.

\section{Conclusion}
Significant irreversible heating with ultrahigh magnetic fields was observed in the $\alpha$-$\theta$ phase transition of solid oxygen.
The occurrence of the first-order phase transition was detected with a simple setup for temperature measurement.
A discontinuous temperature increase from 13 to 37 K, amounting to $\Delta Q=700$ J/mol, was observed immediately after applying a pulsed magnetic field of 137 T.
The huge amount of irreversible heating was due to the fast sweep speed of the STC.
Because of this heating, the phase transformation occurred as $\alpha \rightarrow \theta$ in the up sweep while $\theta \rightarrow \beta$ in the down sweep, which was confirmed by magnetization measurement and optical spectroscopy.
The correspondence between the temperature and the other measurements verified the reliability of the temperature  measurement with a time resolution of 0.34 s.
In the control experiment where the phase transition was absent, no heating was observed.
Therefore, the observed heating effect is attributed to the $\alpha$-$\theta$ phase transition of solid oxygen.
We found that the observed heating is more than twice the magnetic hysteresis loss, $\mu_0 \oint B dM=330$ J/mol.
We indicated the effect of the structural transformation as the origin of this discrepancy.
Frictional motion of the domains at the structural transformation will contribute to the irreversible heating in addition to the magnetic hysteresis loss.

We emphasize that this is the first thermometric measurement conducted with ultrahigh magnetic fields of over 100 T.
When ultrahigh magnetic fields are generated by a destructive technique, many experimental difficulties are imposed.
In the measurement introduced in this paper, all the difficulties are avoided by relinquishing the information during the pulse field generation.
The advantage of this measurement is its simplicity: the temperature of the sample is simply measured with a good thermal contact.
This measurement can be combined with most physical property measurements without interference if the sample volume is sufficiently large for the adiabatic condition.
This can be a powerful subsidiary probe for STC experiment which requires many resources and considerable time for repeated measurement.

\section*{Acknowledgments}
We thank Y. Kohama for helpful discussions.
TN was supported by Japan Society for the Promotion of Science through the Program for Leading Graduate Schools (MERIT) and a Grant-in-Aid for JSPS Fellows.


\begin{thebibliography}{99}
\bibitem{relaxation_t}
Here, the sample shape is approximated by a cylinder with a diameter of 2.4 mm.
The following parameters for the $\beta$ phase of solid oxygen at 30 K are used \cite{04PR_Freiman}: 
thermal conductivity 0.35 W/m$\cdot$K, heat capacity 30 J/mol$\cdot$K, molar volume 21$\times10^{-6}$ m$^3$/mol, and thermal diffusivity 0.25$\times10^{-6}$ m$^2$/s.
\end{thebibliography}
\end{document}